\def\aa{A\&A}               %Astronomy & Astrophysics%
\def\anj{AJ}                %Astronomical Journal%
\def\apj{ApJ}               %Astrophysical Journal%
\def\baas{BAAS}             %Bulletin of the American A.S.%
\def\mn{MNRAS}              %Monthly Notices of the Royal...%
\def\nat{Nature}            %Nature%
\def\rev{ARA\&A}            %Annual Review of Astronomy & Astrophysics%
\documentclass{aa}

\begin{document}

\thesaurus{3(11.01.2; 11.09.1 J1835+620; 11.10.1; 13.18.1)}

\title{Restarting activity in the giant radio galaxy J1835+620}

\author{L. Lara\inst{1}\thanks{Visiting Astronomer, German-Spanish
Astronomical Center, Calar Alto, operated by the Max-Planck-Institut
fur Astronomie jointly with the Spanish National Commission for
Astronomy} \and 
I. M\'arquez\inst{1\star} \and
W.D. Cotton\inst{2} \and
L. Feretti\inst{3} \and
G. Giovannini\inst{3,4} 
\and
J.M. Marcaide\inst{5} \and
T. Venturi\inst{3}}

\offprints{L. Lara}

\institute{Instituto de Astrof\'{\i}sica de Andaluc\'{\i}a (CSIC),
Apdo. 3004, 18080 Granada (Spain)
\and 
National Radio Astronomy Observatory, 520 Edgemont Road, Charlottesville, 
VA 22903-2475 (USA)
\and
Istituto di Radioastronomia (CNR), via P. Gobetti 101, 40129 Bologna (Italy)
\and
Dipartamento di Fisica, Universit\'a di Bologna, via B. Pichat 6/2,
40127 Bologna (Italy)
\and
Departamento de Astronom\'{\i}a, Universitat de Val\`encia, 46100 Burjassot - 
Spain
} 
\date{Received / Accepted}

\authorrunning{Lara et al.}
\titlerunning{Restarting activity in the giant radio galaxy J1835+620}

\maketitle

\begin{abstract}

We present radio and optical observations of the peculiar radio galaxy
J1835+620, a member   of a  new  sample of  large angular   size radio
galaxies   selected from  the  NRAO   VLA   Sky Survey.   Its  optical
counterpart is in a group of at least three galaxies and shows strong,
narrow  emission  lines which allow us    to measure its   redshift as
$z=0.518$.  The derived cosmological  distance places  J1835+620 among
the giant radio galaxies.   The  outstanding  aspect of J1835+620   at
radio wavelengths is the  existence of two symmetric bright components
within  a typical Fanaroff-Riley type II  structure.  We interpret the
source as the result of  two distinct phases of  core activity.  Radio
maps and radio polarization properties are consistent with a dense new
ejection evolving  through an older underlying  jet.  We  suggest that
interaction  with  nearby  galaxies could be the reason for restarting 
the activity in J1835+620.

\end{abstract}

\section{Introduction}

Radio sources  are born,    grow   and  finally\ldots  sleep.     This
evolutionary process can be inferred from the diversity of known radio
sources:  compact symmetric  objects  are  likely  to  be  young radio
sources    (Readhead et al.    \cite{readhead}); Fanaroff-Riley  radio
galaxies of type I  and II (Fanaroff \&  Riley \cite{fanaroff}) may be
considered   ``adult'' sources; finally,   relic sources  are probably
associated with galaxies    which have ceased their   nuclear activity
(Komissarov         \&    Gubanov   \cite{komissarov};      Harris  et
al. \cite{harris}).  If we  believe that  the  activity in  radio-loud
active   galactic nuclei is  the result  of  accretion  onto a compact
massive object, likely a black hole, the life of  a radio source would
be subordinated to the   accretion rate.  The vanishing   of accretion
would  lead  a former radio source   to  a ``dormant''  or hibernation
phase. Such a picture is supported by the increasing number of massive
dark  objects  detected in inactive   galaxies (see Kormendy  \&
Richstone \cite{kormendy}).   However,  interaction  and merging  with
neighboring galaxies can trigger the  activity, and eventually produce
a transition from  a dormant to an  active phase (Stockton \& Mackenty
\cite{stockton};   Bahcall  et  al.   \cite{bahcall}   and  references
therein). 

Under such a  scenario, it should be expected  that a number of  radio
sources with clear evidences of having passed through different phases
during their lifetime  are found.  A  promising candidate is the radio
source 3C338, with a large-scale structure apparently unrelated to the
present nuclear activity (Giovannini et al. \cite{giovannini}).  Other
sources, like 3C219 (Clarke et al.  \cite{clarke2}) or 3C33.1 (Rudnick
\cite{rudnick2}), show  indication     of episodic events     possibly
resulting from an alternation of high and low activity phases. 

In  this paper,  we   present VLA  observations of  the   radio source
J1835+620, made within the frame of the study of a new sample of large
angular size   radio galaxies (Lara et al., in preparation) selected from  
the   NRAO VLA Sky Survey
(NVSS; Condon et al. \cite{nvss}). Optical imaging and spectroscopy of
the host galaxy made at the Calar Alto Observatory are also presented.
No previous  studies  of this  radio source have    been found in  the
literature.  J1835+620 presents, as its  main peculiarity, clear signs
of two distinct phases of nuclear activity. 

\begin{figure*}
\vspace{14cm}
\includegraphics{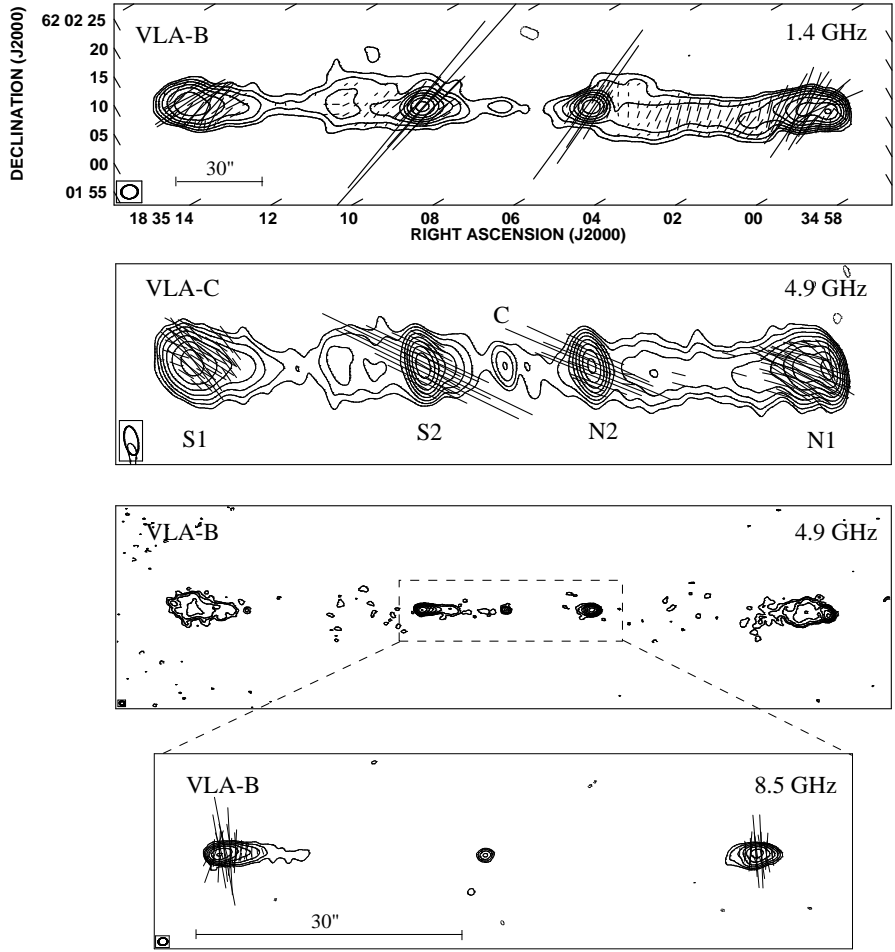}
%\rule{0.4pt}{4cm}% line thickness, height of picture
\caption{VLA total intensity maps of J1835+620 at 1.4, 4.9 and 8.5 GHz, 
rotated clockwise on  the sky  by 60\degr.    Contours are spaced   by
factors of 2 in brightness,  with the lowest at 3  times the rms noise
level.   The vectors    represent  the  polarization   position  angle
(E-vector),   with   length     proportional  to   the    amount    of
polarization. From top  to  bottom, we list  the rms  noise level, the
equivalence of  1\arcsec~  length   in  polarized intensity  and   the
Gaussian beam used  in convolution: rms = 0.2,  0.06, 0.045 and  0.039
mJy  beam$^{-1}$; 1\arcsec~  = 200,  33 and  167  $\mu$Jy beam$^{-1}$;
Beam=   $5\arcsec.8 \times  4\arcsec.7$ P.A.   -25\degr,  $10\arcsec.2
\times    4.7$    P.A.    72\degr,   $1\arcsec.8   \times  1\arcsec.4$
P.A. -43\degr, $1\arcsec \times 0\arcsec.8$ P.A. -24\degr.} 
\label{mapas}
\end{figure*}

\section{Observations and results}

\subsection{Radio observations}

We made continuum observations of J1835+620 with the VLA in its B- and
C-configurations  at 1.4,  4.9 and 8.5  GHz  (see Table \ref{obs}  for
details). The  radio sources 3C286 and/or  3C48 served as primary flux
density calibrators.  The   interferometric phases  of J1835+620  were
calibrated  using the nearby radio  sources J1927+612 (at  1.4 and 4.9
GHz) and J1849+670  (at 8.5 GHz).   The  processes of self-calibration
and  imaging  of the  data  in total  intensity and  polarization were
carried  out  with  the  NRAO    AIPS  package,  following    standard
procedures. Maps  at 4.9 and  8.5 GHz had to  be corrected for primary
beam attenuation;   such effects   were  negligible on  the  1.4   GHz
maps. Polarization maps were corrected for the non-Gaussian-like noise
distribution of the polarized intensity. 

\begin{table}[]
\caption[]{VLA observations of J1835+620}
\label{obs}
\begin{tabular}{ccccr}
\hline
Array  & $\nu$ & $\Delta\nu$ & Duration & Date~~~ \\
       & (MHz) & (MHz)       & (min)    &      \\
\hline
 B      &  1442  & 50  &  8   & 18 Feb 97 \\
 B      &  4860  & 100 &  10  & 18 Feb 97 \\
 B      &  8460  & 100 &  30  &  7 Oct 98 \\
 C      &  1425  &  50 &  10  & 29 Dec 98 \\ 
 C      &  4860  & 100 &  10  & 29 Dec 98 \\
\hline 
\end{tabular}
\end{table}

Radio maps of   J1835+620  are displayed in    Fig.~\ref{mapas}, after
having been rotated clockwise on the  sky by 60\degr.  At first sight,
a very  peculiar radio structure  is  evident: a compact  central core
(labeled  C  in Fig.~\ref{mapas})  and  two radio  lobes   (N1 and S1)
straddling  two  bright  components  (N2  and  S2).  The  total source
angular  size is $3\arcmin.88$,  the arm-ratio not being significantly
different from unity. The  angular distance between components  S2 and
N2 is $1'.07$, with an arm-ratio  of 0.95 (south  to north).  From our
highest  angular resolution  observations, we  determine the following
coordinates for the compact core (J2000): RA $18^h 35^m 10^s.408$; DEC
$62\degr~ 04\arcmin~07\arcsec.42$. 

Although  the structure  of J1835+620 is    very symmetric, there  are
differences between the  two sides  of  the source:  the northern  arm
shows a continuous  and  well collimated  bridge of  emission  joining
components N1 and N2; the southern  arm presents a similar bridge, but
with  a clear discontinuity in  between.  Both bridges  have a similar
spectral index $\alpha_{1.4}^{4.9} = 1.2$ (the spectral index $\alpha$
is defined so that $S\propto \nu^{-\alpha}$). The N1 lobe has a higher
flux density than S1, while N2 is weaker than S2. Finally, the N1 lobe
hosts a compact hot-spot, while no hot-spot of similar characteristics
is observed in the S1 lobe. 

The  radio core presents  signs   of  flux density variability:    our
observations at 4.9 GHz, spanning 20 months,  show a core flux density
decrease from 2.1 mJy  at 1997.13 to  1.5 mJy at 1998.99, a  variation
well above the flux density calibration errors ($\le 10$\%). 

Electric field vectors   with  lengths proportional to  the  polarized
intensity are also displayed in Fig.~\ref{mapas}.  The mean fractional
polarization ($p_m$) reaches 20\% in  N2 and S2,  and about 12\% in N1
and S1, at 1.4  and  8.5 GHz.   However, at  4.9  GHz we  measure only
$p_m=8$\% in N2 and  S2 and $p_m=5$\% in N1  and S1.   Since different
frequency  observations  show clear  signs of Faraday  rotation of the
polarization  vectors, we  made a  pixel  by  pixel evaluation of  the
rotation measure ($RM$) over  the entire source  in  order to  test if
such a drop in  the mean fractional polarization  could be produced by
small   scale   fluctuations in   the   rotation measure distribution.
However,  at each position the derived  $RM$ is, within errors, always
compatible with the mean $RM$ determined from component  N2: $RM = 276
\pm 7$ rad m$^{-2}$ with  an intrinsic position  angle of the electric
vector   of $60\degr$.  As    a consequence,  the   drop  in  the mean
fractional polarization  at 4.9 GHz  can  hardly  be ascribed to  beam
depolarization  or to  finer  structure  in the  $RM$ distribution;  its
origin  remains unclear to    us.  Unfortunately, VLA B-configuration
observations at 4.9 GHz could not be  properly calibrated in polarized
intensity due to insufficient coverage of the paralactic angle. 

Flux  densities (S), spectral indexes  ($\alpha$)  and mean fractional
polarizations ($p_m$) of components N1, N2, C, S1 and S2 are displayed
in Table~\ref{comp}  for  the different  frequencies  of  observation.
Although  our maps might   be    affected  by missing  flux    density
(marginally at 1.4 GHz and 4.9 GHz; more  importantly at 8.4 GHz since
the  C-array was not used), we  note  that this mainly affects regions
not considered in  the flux density and  spectral index measurements in
Table~\ref{comp}.   As  a test, we compared  the  flux  density of the
different components measured from the VLA-C and the VLA-B maps at 4.9
GHz, obtaining compatible results.  Moreover, the spectral  index does
not show the remarkable steepening  which should be  expected   if 
significant flux density were unrecovered at 8.4 GHz in these components.

\begin{figure*}
\vspace{14cm}
\includegraphics{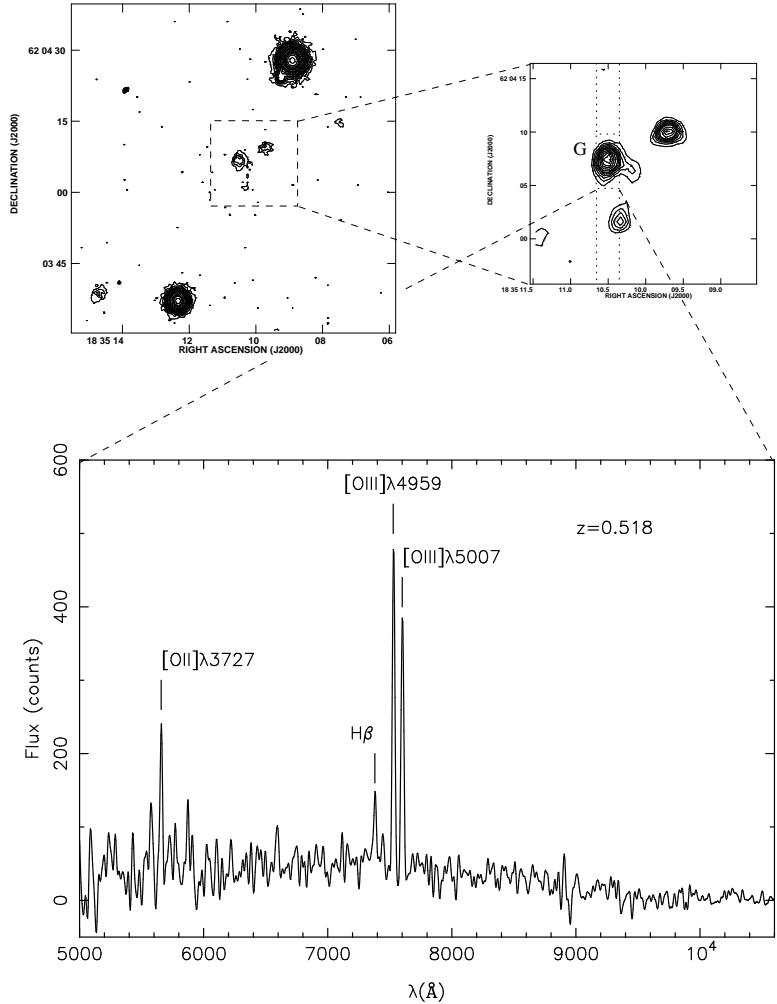} 
%\rule{0.4pt}{4cm}% line thickness, height of picture
\caption{Optical observations of J1835+620: The upper-left panel shows 
an R-filter image  centered at the  position of the  radio core.   The
upper-right  panel  shows  a detail  of the  optical  image after  the
application of deconvolution algorithms.  The host galaxy of J1835+620
is  labeled with  the letter ``G''.   The dotted  lines represent  the
orientation  (N-S)  and width   ($2\arcsec$) of  the  slit during  the
spectroscopic observations, and the  region selected ($5\arcsec.3$  in
diameter) to construct the 1-D spectrum shown at the bottom panel.} 
\label{optico}
\end{figure*}

\begin{table*}[t]
\caption[]{Radio components in J1835+620}
\label{comp}
\begin{tabular}{ccccccccc}
\hline
Component  & S$_{1.4}$ & S$_{4.9}$ & S$_{8.5}$ & $\alpha_{1.4}^{4.9}$ 
& $\alpha_{4.9}^{8.5}$ & $p_m^{1.4}$ & $p_m^{4.9}$ & $p_m^{8.5}$  \\
       & (mJy) & (mJy) & (mJy) &  &  & (\%) & (\%) & (\%) \\
\hline
 N1  & 227 &  79   &  52 & 0.87 & 0.73    & 12   &  5   &  13   \\
 N2  &  90 &  33   &  19 & 0.82 & 0.88    & 19   &  8   &  20   \\
 C   & 2.3 &1.5$^a$& 1.2 & 0.07 & 0.40$^b$&$\le$2&$\le$3&$\le$2 \\
 S2  & 139 &  51   &  29 & 0.82 & 0.91    & 20   &  8   &  18   \\ 
 S1  & 149 &  53   &  28 & 0.85 & 0.97    & 15   &  5   &   9   \\ 
\hline
Total& 746 & 248   & 135 & 0.91 & 1.05    & --   & --   &  --   \\
\hline 
\multicolumn{9}{l}
{\parbox{10cm}{\footnotesize{$^a$ Flux density at epoch 1998.99\\
$^b$ Derived from non-simultaneous epochs: 1998.77 and 1998.99}}}
\end{tabular}
\end{table*}

\subsection{Optical imaging and spectroscopy}

We made optical observations of  J1835+620 centered on the position of
the radio core on the  night of 1998 October 30  at the 2.2m telescope
in  Calar  Alto  (Spain).    We used the    Calar  Alto Faint   Object
Spectrograph (CAFOS) equipped with a SITe $2048\times2048$ CCD.  CAFOS
allows  direct  imaging  and spectroscopy,  with  a  spatial scale  of
$0''.53$  per pixel.  We  used a  medium resolution grism (200\AA/mm),
sensitive to the wavelength range of 4000  to 8500 \AA~, that provides
a  spectral  resolution of  4.47  \AA/pixel.  The  image and  spectrum
obtained are shown in Fig.~\ref{optico}. 

To identify and position the optical counterpart of the radio core, we
first  obtained a single 300s exposure  image in the R-band, finding a
galaxy  coincident  with the radio  core.   Then,  with the long  slit
configuration (with a $2''$  wide slit placed north-south) we obtained
a spectrum of 2400s in  two equal exposures,  to be added up
in order  to reject  the cosmic rays.   We measured  a seeing  of FWHM
$1\arcsec.5$   using the foreground stars in    the R image.  The data
reduction  and   calibration  were performed  following   the standard
procedures with  the  IRAF\footnote{IRAF is  the  Image Reduction  and
Analysis Facility made available  to the astronomical community by the
National  Optical Astronomy Observatories,   which are operated by the
Association of Universities for  Research  in Astronomy (AURA),  Inc.,
under contract with the U.S.  National Science Foundation.}  software,
involving dark and flat field corrections.  Wavelength calibration was
carried out   using exposures  of mercury-helium-rubidium lamps  taken
before  and  after  the  target exposure.   No   flux calibration  was
attempted as  the atmospheric  conditions  were not  photometric.   We
finally  extracted and summed 10 spectra  centered with respect to the
core position, corresponding to an extension of $5\arcsec.3$. The slit
and the extracted region are shown in Fig.~\ref{optico}. 

The 1-D spectrum  shows a very  weak continuum (barely  visible in the
2-D   spectrum),  with  prominent  H$\beta$,  [OII] and  [OIII] narrow
emission lines, from which we determine a redshift z=0.518. 

In Fig.~\ref{optico}  we   also show an  enlargement of    the central
$20\arcsec \times 20\arcsec$  of  the image deconvolved  using  Lucy's
algorithm within IRAF, with  a  resulting  FWHM of $0\arcsec.8$.    In
addition to the radio  core  host galaxy, elongated  along north-south
direction and  with  an extension to the  west,  we clearly detect the
presence of two other galaxy-like objects in  the very close vicinity,
one at $5\arcsec.9$ to  the west and the  other at $5\arcsec.4$ to the
south. 

The estimated redshift of   J1835+620  implies a total radio    source
length of  1.12  Mpc ($1\arcsec$  corresponds  to 4.81  kpc in  linear
distance, assuming H$_{0}=75$ km s$^{-1}$ Mpc$^{-1}$ and q$_{0}=0.5$),
placing J1835+620 among    the  giant radio galaxies.   The   distance
implied  between  components N2 and  S2  is  309  kpc. The two nearby 
galaxies are 
at a projected distance of 28.4 kpc (west galaxy) and 
26.0 kpc (south galaxy) from the radio core host galaxy.

\section{Discussion}

The most distinguishing  aspect in J1835+620  is the  existence in its
radio  structure of  two  symmetric  and  bright  components within  a
typical FR II radio source.  This fact  prompts us to invoke scenarios
of  restarting or   recurrent   activity  in active galactic    nuclei
(Christiansen  \cite{christiansen}).   At  first  sight,  the apparent
source  symmetry  does not favor  models  in which ejections  from the
nucleus  occur on only one   side at a time (``flip-flop''  mechanism,
e.g.   Rudnick \&   Edgar  \cite{rudnick1}).  Instead,   ejections  in
J1835+620 appear to be simultaneous, supporting scenarios in which the
activity of the central  core  is  alternately  switched on and   off.
Accordingly, the morphology of J1835+620 could well be the consequence
of two periods  of activity in  the  core, separated by a  ``dormant''
phase.  We  note  that restarting   activity resembles   the  commonly
observed ejection  of moving components  in  parsec-scale jets.   But
while such  components are most  possibly  due to  the propagation  of
shock waves along continuous jets  (Marscher \& Gear \cite{marscher}),
at megaparsec  scales much more  dramatic physical conditions would be
required. 

Clarke \&  Burns (\cite{clarke1}) have  made 2-D numerical simulations
of restarting jets evolving through a medium already ``cooked'' by the
original  ejection.  They  find  a number of  properties  which should
distinguish new ejections from original ones.  First, the new ejection
is always denser  than the surrounding medium if  the original  jet is
``under-dense'' relative to  the intergalactic medium  (IGM).  This is
because the new jet evolves in a hot  and rarefied medium processed by
the   original jet.  As   a consequence,  the   advance  speed of  the
restarted jet is greater than that of the original one, while the Mach
number is lower.  Second, the bow shock  and the terminal Mach disk in
the new jet are of comparable strength. Therefore,  if no bow shock is
observed, the emission from  the  new jet should  be  weak.   However,
recent  3-D   simulations show a  well-defined  bow  shock leading the
restarting  jet   (Clarke  \cite{clarke3}).   Finally,  the  numerical
simulations show that without the momentum flux of the jet an existing
hot-spot in a lobe  would expand rapidly, decreasing dramatically its
level of synchrotron emission  in a time short  compared to the on-off
duty cycle. 

We wish to understand if the restarting scenario really describes what
is  happening in J1835+620. If so, components  N2 and  S2 would represent the
evolution of a  new ejection, and this  situation must leave traces  in
the new components which we should identify. 

\subsection{Radio spectral properties}

The existence of a  hot-spot  in  N1 poses   difficulties  for  the
assumption of  a core whose  activity has been completely switched off
and resumed some time  later.  If the N1   lobe were the result of  an
already exhausted phase   of   activity, the presence  of   a  compact
hot-spot would be difficult to  explain without a continuous supply of
momentum flux  (Clarke  \& Burns  \cite{clarke1}).  The  most  natural
explanation is that the lobes are still supplied through an underlying
jet, while N2 and S2 are the result of a dramatic increase in the core
activity. 

This scenario is  supported by the comparison of  the spectral ages of
components N1  and S1  with those of  N2 and  S2: considering  that no
significant differences  in the   spectral  break frequencies  can  be
inferred  from  our data, we   obtain  a synchrotron   spectral age of
$3.5\times  10^6$  yr for components   N1  and S1, and  of  $2.3\times
10^{6}$ yr for components  N2 and S2.  In this calculation we assume a
break frequency of 20  GHz  consistent with our observations,  minimum
energy conditions and take into account the Inverse Compton losses due
to the interaction with  the microwave background.  The  estimated age
of the ``new'' components is in agreement with a dynamical age derived
from an expansion velocity of 0.20-0.25 c;  the synchrotron age of the
``old'' components requires  an unplausible expansion velocity of  0.5
c.  Therefore, spectral  ages  are  not  consistent  with an exhausted
primary phase of activity, implying that N1  and S1 are still supplied
by fresh   particles.   An  underlying jet must   exist  in J1835+620,
through which  the new components,  N2 and  S2, evolve.  This scenario
would  explain  the  absence   of a bow    shock  preceding these  new
components (Clarke \cite{clarke3}), since N2  and  S2  would propagate 
through an already existing jet. 

\subsection{Magnetic field}

\begin{figure}
\vspace{6.5cm}
\includegraphics{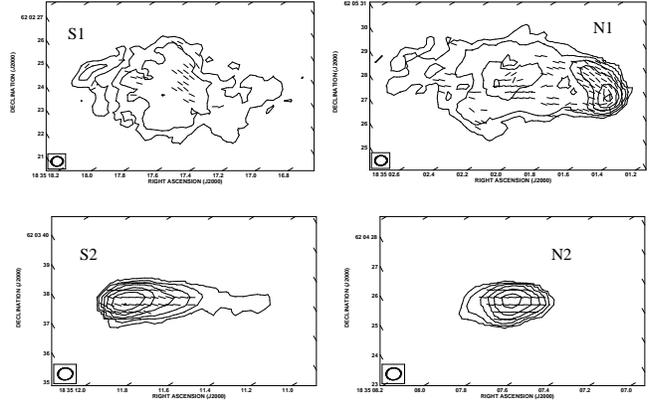}
%\rule{0.4pt}{4cm}% line thickness, height of picture
\caption{Total intensity maps of the S1, N1, S2 and N2 components  
of J1835+620 at 8.5 GHz, with vectors superimposed showing the intrinsic 
orientation of the projected magnetic field. Vector lengths have been 
normalized to unity.}
\label{magne}
\end{figure}

The study of the  magnetic field configuration is  a powerful tool  to
test and constrain the restarting  model requirements. Fig.~\ref{magne} 
displays the
orientation  of   the  projected magnetic  field   in  J1835+620, once
corrected for Faraday rotation.  Components N2 and S2 have a magnetic
field parallel to the jet axis. On the  other hand, component N1 shows
the typical structure of a  cocoon expanding in  the IGM medium,  with
the magnetic field  compressed at the  hot-spot, and parallel to  the
cocoon  edges.  Component  S1,  however,  does not  show   evidence of
compression,  compatible with the absence of   a bright hot-spot.  The
fact  that the  magnetic  field is  parallel in N2   and S2 means that
little  or no compression  affects the
evolution   of  the  new  components, in  favor  of an over-dense  ejection
with respect to the underlying jet. That would indicate that the two activity 
phases in the core of J1835+620 produce jets with different properties.

The origin  of the differences in flux density between components N2
and S2 remains unclear  to us.  
They could  be due to  intrinsic differences in the ejection,
or to differences in the medium of propagation, or to a combination of these 
with orientation related effects. 

The uniformity of  the $RM$ over J1835+620  places, most  plausibly, the
ionized screen producing  Faraday rotation  outside the radio  source.
Since the expected $RM$ of galactic origin at  the position of J1835+620
is much smaller than measured (Simard-Normandin \& Kronberg 
\cite{normandin}), the origin of  Faraday  rotation in J1835+62  would
most  likely   be associated with  the  existence  of an  ionized halo
surrounding  the   galaxy.   We do  not  find  evidence  of asymmetric
depolarization in this radio source  (the so called Laing - Garrington
effect,  Laing \cite{laing};  Garrington  et  al.  \cite{garrington}).
Assuming the existence of a halo, this  fact could be explained if the
source  main axis were  lying close  to  the plane of   the sky.  This
orientation, consistent with the  morphological symmetry of  the radio
source,   would  explain the  very low  flux  density  of the core (if
unification     schemes   of   radio-loud     active  galactic  nuclei
(e.g. Antonucci  \cite{antonucci}) are  correct), and favors J1835+620
being a giant radio galaxy.

\subsection{Restarting activity}

The reasons  for restarting, or more properly, amplifying  the   activity 
in  the core  of J1835+620
remain  a mystery.  It is  already known  that  galaxy interaction and
merging   can   trigger   nuclear  activity    (Stockton  \&  Mackenty
\cite{stockton}; Bahcall et   al. \cite{bahcall}).  Our optical  image
shows that the  radio  core of  J1835+620 resides  in  a  galaxy which
belongs to  a group of at  least three galaxies showing marginal signs
of    mutual interaction. It  is     thus  tempting to  suggest   that
interactions in this group of  galaxies  could be associated with  the
different stages of radio activity in J1835+620. Deeper optical images
and spectroscopy  of the  other two members   of the group  might shed
light into this conjecture. 

Other interpretations for  J1835+620 could in principle be considered.
Clarke et al. (\cite{clarke2})    proposed, and applied on  3C219,   a
``passive   magnetic  field''   model   in which    the  observed  jet
interruptions  would be a consequence  of a transition from a dominant
axial  magnetic field  to a  dominant toroidal field  occurring at the
position where  a force   balance between a  continuous  jet  and  the
surrounding  medium  is attained.   However,  we  do  not  observe the
predicted  magnetic    field      transitions   in    J1835+620   (see
Fig~\ref{mapas}), rendering this model implausible. On the other hand,
arguments  based on recollimation  shocks to explain components N2 and
S2 have difficulty  in explaining the  extraordinary  symmetry of this
source.

\section{Conclusions}

Radio and optical observations of J1835+620 have been
presented. The observations  were made in the frame  of the study of a
new sample of large angular size radio sources selected from the NVSS.
The R-band optical image shows a galaxy coincident with the radio core
position.  There  are other two galaxies  closely  seen in projection.
The spectrum of the core host galaxy shows a very weak continuum,
but prominent narrow emission lines from which we determine a redshift
$z=0.518$.   Consequently,  with  an  angular  radio size of $3\arcmin.88$,
J1835+620 is classified as a giant radio galaxy. 
 
J1835+620  shows a very  peculiar radio structure,  with two symmetric
bright components within  a typical Fanaroff-Riley type II  structure.
Given the  extraordinary symmetry of the  radio  source (S1 to  N1 
arm-ratio = 1; S2  to N2 arm-ratio = 0.95), we  interpret it using the
model of restarting activity  in radio galaxies.  Two distinct  phases
of core activity would be responsible for the  observed  morphology.  

Observational properties  have  been
compared with predictions from 2-D numerical  simulations by Clarke \&
Burns (\cite{clarke1}). The existence of a hot-spot in N1 together with
spectral aging arguments indicate that N1 and S1 are still supplied
by fresh particles, implying {\em i)} the existence of an underlying jet
connecting the core with the outer components and {\em ii)} that the activity 
in J1835+620 did not
stop completely. In consequence, the new components N2 and S2 would represent
the result of a new ejection propagating through the primary underlying jet.   
The  parallel magnetic fields  of components N2
and  S2 are consistent  with a ``second-phase'' jet which is overdense with 
respect to the ``first-phase'' one.

Faraday rotation of  the polarization vectors is  observed, with a  $RM$
distributed rather uniformly all over  the source. We suggest that the
rotation is produced by an ionized  halo surrounding the radio source.
Asymmetric depolarization is not observed, which could be explained if
the main source axis were oriented approximately parallel to the plane
of the  sky. This orientation would  also be consistent with the large
size of the source and with the low flux density of the core. 

The conditions under which  a galaxy can change  its degree of nuclear
activity remain  unknown.   Interaction   between galaxies   is  often
invoked  as a reason for  triggering  nuclear activity.  We find hints
that interaction of  the galaxy  hosting  the radio  core with  nearby
galaxies might be  taking place,  but  this point has to  be confirmed
through deeper optical imaging and  spectroscopy of the members of the
group of galaxies.

\begin{acknowledgements}

We thank the referee Dr. J. Condon for helpful and constructive comments to 
the paper. This  research  is supported in  part  by  the Spanish DGICYT
(PB97-1164 and PB93-0139). GG acknowledges the Italian Ministry for 
University and Research (MURST) for financial support under grant 
Cofin98-02-32. The National Radio Astronomy Observatory is a 
facility of the National Science Foundation  operated under cooperative  
agreement by Associated Universities, Inc. 

\end{acknowledgements}

\end{document}